\documentclass[12pt]{article}
\usepackage[mathscr]{eucal}
\usepackage{dutchcal}
\usepackage{bm}
\usepackage{bbold}
\usepackage{caption}
\usepackage{multirow}
\usepackage{amsmath}
\usepackage{textgreek}
\usepackage[greek,english]{babel}
\usepackage{xcolor}
\usepackage{graphicx,subfigure}
\usepackage{empheq}
\usepackage{amsfonts}
\usepackage{amssymb}
\usepackage{dsfont}
\usepackage{ytableau}
\usepackage{overpic}
\usepackage{hyperref} 
\usepackage{makeidx}
\usepackage{empheq}

\newcommand{\be}{\begin{eqnarray}}
\newcommand{\ee}{\end{eqnarray}}
\newcommand{\bc}{\begin{center}}
\newcommand{\ec}{\end{center}}
\newcommand{\nn}{\nonumber \\}
\newcommand{\lb}{\label}
\newcommand{\pd}{\partial}
\newcommand{\p}[1]{(\ref{#1})}

\def\theequation{\arabic{section}.\arabic{equation}}

\begin{document}

\begin{titlepage}

\vspace*{0.2cm}

\begin{center}

{\LARGE\bf $QCD_2$: `t Hooft Equation vs.  Relativistic Schr\"odinger Equation}

\vspace{2cm}

{\Large Andrei Smilga}\footnote{Smilga@subatech.in2p3.fr} \\

\vspace{0.5cm}

{\it SUBATECH, Universit\'e de Nantes, 
 4 rue Alfred Kastler, BP 20722, Nantes  44307, France.}

\end{center}

\vspace{0.2cm} \vskip 0.6truecm \nopagebreak

   \begin{abstract}
We study two-dimensional QCD with a quark of mass $m$ in the fundamental representation, in the 't Hooft limit ($N \to \infty$, $g^2 \to 0$, with $g^2N$ fixed). The spectrum contains an infinite tower of meson states. Their masses can be found by solving the corresponding integral Bethe--Salpeter equation. A simpler alternative is to solve the Schr\"odinger equation for the relativistic Hamiltonian
\be
\lb{Salp}
\hat H = 2\sqrt{\hat p^2 + m^2} + \sigma |x|
\ee
with the string tension  $\sigma = g^2N/4$. We compare the results obtained by these two methods. In the leading semiclassical approximation, they coincide. However, the preasymptotic WKB corrections to the masses differ. These corrections are important for the ground state, but already for $n=3$ the relative difference between the two predictions is no more than 2.5\% and decreases rapidly as $n$ increases.
\end{abstract}

\end{titlepage}

\section{Introduction}

\setcounter{equation}0

In a seminal paper written half a century ago \cite{Hooft}, \ 't Hooft found the particle spectrum in 2-dimensional quantum chromodynamics with $SU(N)$ gauge group in the limit $N \to \infty$. In two dimensions, there is no gluon self-interaction (its absence is manifest if one imposes the  light-cone gauge), so there are no glueballs. In the infinite $N$ limit, there are no baryons either and the only states are {\it mesons} --- bound states of a quark and an antiquark: the production of extra $q \bar q$ pairs is also suppressed. Consequently, meson states are absolutely stable in this limit. At large but finite $N$, the widths are not zero, but they are suppressed $\propto 1/N$.

't Hooft determined the spectrum by solving the Bethe--Salpeter equation for the vertices relating the bound states to their quark and antiquark constituents.

Let us first specify our notation. It differs somewhat from that used in 't Hooft's paper.

The Lagrangian of the theory reads

\be
\lb{QCD2}
{\cal L} \ =\ \frac 1{g^2} {\rm Tr} \{\hat F_{01}^2 \} + i \bar \psi \gamma^\mu (\pd_\mu - i \hat A_\mu) \psi -m \bar \psi \psi,
\ee
where $\hat A_{\mu = 0,1} = A^a_{\mu = 0,1} t^a \in su(N)$
and $\hat F_{01} = \partial_0 \hat A_1 - \partial_1 \hat A_0 - i [ \hat A_0,\hat A_1]$. Introducing light-cone variables, $\hat A_\pm = (\hat A_0 \pm \hat A_1)/\sqrt{2}$ and similarly for $\pd_\pm$, and imposing the gauge $\hat A_- = 0$, the field density reduces to 
$\hat F_{01} = \pd_- \hat A_+$. The fermion $\psi$ (call it ``quark") belongs to the fundamental representation of $SU(N)$.

 For the coupling constant $g^2$ in Eq. \p{QCD2}, we have used  the standard convention,  which is twice as large as 't Hooft's convention:
$$
g^2_{\rm standard} \ =\ 2 g^2_{\rm Hooft}.
$$

In the infinite momentum frame, the vertex of dissociation of a meson state into its constituents reduces to a simple  function $\phi(r)$, with $r \in (0,1)$ having the meaning of the fraction of the large meson momentum carried by the quark.\footnote{'t Hooft's notation for this variable was $x$, but we have chosen another letter not to confuse it with $x$ in \p{Salp}.} Correspondingly, the antiquark carries the fraction $1-r$. The Bethe--Salpeter equation reduces to the following relatively simple integral equation for $\phi(r)$. It can be written in the form
\be
\lb{eq-Hooft}
 M^2  \phi(r)  = \ \frac { m^2}{r(1-r)}  \phi(r)   - \frac {g^2N}{2\pi} P \int_0^1 \frac {\phi(t)} {(t-r)^2} dt, 
\ee
where $m$  is the renormalized quark mass, $ m^2 = m_0^2 - g^2N/(2\pi)$,  $M$ is the mass of the  meson bound state and the boundary conditions $\phi(0) = \phi(1) =0$ should be imposed. The singularity
at $t=r$ is treated by the {\it Hadamard finite part} prescription: \footnote{For example, 
$$
P \int_{-1}^1 \frac {dt}{t^2} \ =\ -2.
$$
Do not confuse Hadamard's finite part with  Cauchy's principal value. The latter is not defined for the integral in the right-hand side of Eq. \p{eq-Hooft}.}

\be
P \int_a^b dt\, f(t)  \equiv  \int_a^b dt\, \tilde P f(t) \ =\ \lim_{\epsilon \to 0} \, \frac 12 \left[ \int_a^b dt \,f(t+i\epsilon) + 
\int_a^b dt\, f(t-i\epsilon)   \right]. 
\ee

  There are infinitely many solutions for $\phi(r)$ and $M$. In general, this equation can be solved only numerically.  But the asymptotics for  high excitation levels can also be evaluated analytically. 't Hooft did so for the excitations with  masses $M_n$ much exceeding both relevant  scales: $M_n^2 \gg  m^2$ and $M_n^2 \gg g^2N$. In such a case,  
 \be
\lb{2D-spectrum}
M_n^2 \ \approx \ \frac {\pi g^2N}2 n.
\ee
 This behavior resembles Regge trajectories in the hadron spectrum in the real 4-dimensional world.

It was noticed in Ref. \cite{ja} (see also Refs. \cite{Bars1, Bars2} where similar semiclassical formulas for the  meson spectrum   were derived in the string  framework)
that the law \p{2D-spectrum} may be derived in a very simple way as a solution of the ultrarelativistic quantum problem for the  potential  $V(x) = \sigma |x|$ in semiclassical approximation. 
This approximation also works for massive quarks. In particular, in the nonrelativistic case, $m^2 \gg g^2 N$, and for large excitation numbers $n$, the spectrum obtained from the solution of Eq. \p{eq-Hooft} reads
\be
\lb{mu-nerel-WKB}
M_n \ \approx  \ 2m  + \left( \frac{3\pi g^2 N n}{16\sqrt{m}}  \right)^{2/3}.
 \ee
The same result follows from the solution of the Schr\"odinger equation with the nonrelativistic Hamiltonian written in the 
center-of-mass system: 
 \be
\hat H \ = \ \frac {\hat p^2}m + \sigma |x| \,, 
 \ee
where $|x|$ is  the distance between the quarks related by the string with the tension\footnote{In two dimensions, the Coulomb energy of a pair of Abelian charges separated by the distance $R$ is 
$$
E^{\rm Ab}_{\rm Coulomb} \ =\ \frac {e^2R}2.
$$
In the non-Abelian case, we have to replace $e^2$ by $g^2 c_F$ which is $g^2N/2$ for large $N$.}
$\sigma = g^2N/4$.

In the leading WKB approximation, the spectra obtained from the solution of the integral equation \p{eq-Hooft} and the differential Schr\"odinger equation with the Hamiltonian \p{Salp}  coincide for any quark mass. This agreement has a natural explanation.

\section{Classical equivalence}
\setcounter{equation}0

 The following theorem can be proved: \cite{Yulia}:

{\it The classical counterpart of the relativistic Hamiltonian \p{Salp} and the classical counterpart of the quantum 't Hooft Hamiltonian in the  RHS of Eq. \p{eq-Hooft} are related by a canonical transformation.}

We first prove the following lemma: {\it The classical counterpart of the 't Hooft Hamiltonian reads:
\be
\lb{H-Hooft-class}
H(r,y) \ =\ \frac { m^2}{r(1-r)} + \frac {g^2N}2 |y|,
\ee
with $(r, y)$ being canonically conjugated variables.}

Heuristically, the factor $|y|$ in the second term is the Fourier transform of the kernel $1/(r-t)^2$ in Eq. \p{eq-Hooft}.
Let us do it carefully.

The specific feature of the Fourier transform in this problem is that $ y \in (-\infty, \infty)$ as usual, but the variables $r,t$ belong to a finite interval, $r,t \in (0,1)$. The direct and inverse transformations are defined as follows:
\be
\lb{Fourier}
\phi(t) \ = \ \frac 1{2\pi} \int_{-\infty}^\infty  dy\, e^{iyt} \tilde \phi(y), \qquad \qquad
\tilde \phi(y) \ = \ \int_0^1  dt \, e^{-iyt}  \phi(t).
\ee
We use the identity
\be
\tilde P \frac 1{t^2} \ =\ -\frac 12 \int_{-\infty}^\infty dy |y| e^{iyt}.
\ee
Indeed, 
$$ I =  \frac 12 \int_{-\infty}^\infty dy  |y| e^{iyt} \ =\ \frac 12 \int_0^\infty y dy ( e^{iyt} +  e^{-iyt} ). $$
Regularizing the integral gives
$$
I \ =\ \frac 12 \lim_{\epsilon \to 0} \, \left[  \int_0^\infty y dy \,  e^{(it-\epsilon) y}  +  \int_0^\infty y dy \, e^{-(it+\epsilon) y} \right]  = - \lim_{\epsilon \to 0}\,  \frac 12 \left[\frac 1{(t + i\epsilon)^2} +  \frac 1{(t - i\epsilon)^2} \right] \ =\ - \tilde P \frac 1{t^2}
$$
Then the right-hand side of Eq. \p{eq-Hooft} acquires the form
\be
\lb{Ham-Hooft}
\hat H^{\rm Hooft} \phi(r) \ =\ \frac { m^2 }{r(1-r)}\phi(r) \ +\ \frac {g^2N}{4\pi} \int_0^1 dt \, \phi(t) \int_{-\infty}^\infty dy |y| e^{iy(r-t)}.
\ee
Integrating over $t$ with the help of \p{Fourier}, we represent the second term as
$$
\frac{g^2 N}2 \times \frac 1{2\pi} \int_{-\infty}^\infty dy |y| \tilde \phi(y) \, e^{iyr}.
$$
Looking at Eq. \p{Fourier} again, we see that the second term in \p{Ham-Hooft} is equal to the Fourier image of $|y| \tilde \phi(y)$ multiplied by $\frac{g^2N}2$. This gives the contribution $\frac {g^2N |y|}2 $ in the classical counterpart (in a more mathematical language, the {\it symbol}) of the quantum operator \p{Ham-Hooft}.

\vspace{1mm}
The canonical transformation reducing \p{H-Hooft-class} to
$$ {\cal M}^2 \ =\ \left(2\sqrt{p^2 +  m^2} + \frac {g^2 N}4 |x|  \right)^2$$
reads
\be
\lb{canon}
r \ &=& \ \frac 12 + \frac p {\cal M}, \nn
y \ &=& \ \frac x2 \cdot \frac {{\cal M}^2 ({\cal M} + 2\sqrt{p^2 +  m^2})}{{\cal M}^2 - 4p^2}.
\ee
The property
$$
\{r, y\} \ =\ \frac {\pd r}{\pd p} \frac {\pd y}{\pd x} - \frac {\pd r}{\pd x} \frac {\pd y}{\pd p} \ =\ 1
$$ 
holds.

It is clear now why the leading WKB asymptotics  for the quantum problems \p{Salp} and \p{eq-Hooft} coincide. If the classical theories are equivalent, also quantum theories should stay equivalent in the leading semiclassical approximation.

\section{Nonrelativistic spectrum}
\setcounter{equation}0

In general, the exact spectra of the two quantum problems may be different. And they are --- we will see it a little bit later. But in the nonrelativistic regime, where $g \ll m$ and the meson masses $M_n$ are close to $2m$ with small corrections, the exact spectra coincide. This follows from the well-known fact that in the nonrelativistic limit the Bethe-Salpeter equation reduces to the Schr\"odinger one.

The exact solution of the Schr\"odinger problem with the linear potential $V(x) = \sigma x$ is expressed in terms of the {\rm Ai}ry functions.
In our case, the potential is $V(x) = \sigma |x|$. It is even in $x$, so the spectrum splits into two sectors: with even and odd wave functions.

\begin{figure}
\bc
\includegraphics[width=.6\textwidth]{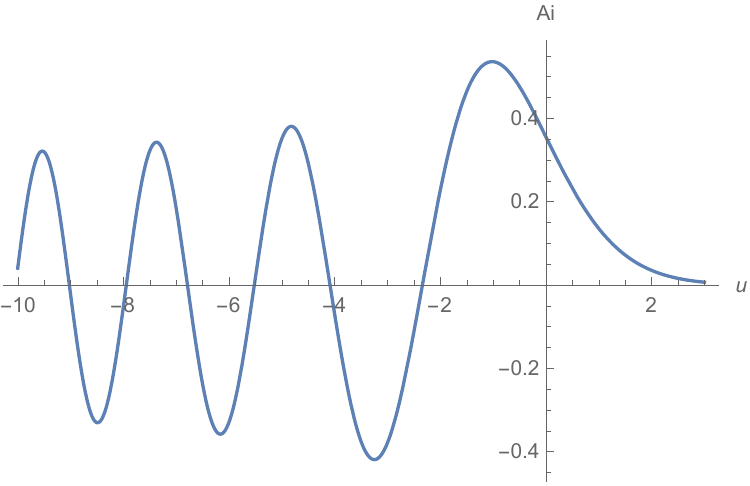}
\ec
        \caption{Airy function}
\label{Airy-fig}
\end{figure}

The odd wave functions vanish at zero, and we can consider the Schr\"odinger problem, 
\be 
\lb{Schr}
\left[ -\frac 1{ m^2} \frac {\pd^2}{\pd x^2} + \sigma x \right] \Psi(x) \ = \ E\Psi(x),
 \ee 
on the half-line $(0, \infty)$ with wave functions vanishing at both ends.  A generic solution to this equation is 
\be
\lb{Airy}
\psi(x) \ =\ {\rm Ai}\left[\mu\left(x - \frac E\sigma \right)   \right],
  \ee
where
 \be
\lb{mu}
\mu = ( m\sigma)^{1/3}
 \ee
 and ${\rm Ai}(u)$ satisfies the equation
\be
\lb{Airy-eqn}
 {\rm Ai}''(u) - u \,{\rm Ai}(u) \ =\ 0.
 \ee
The function  \p{Airy} decays at infinity for any $E$. The condition $\Psi(0) = 0$ makes the spectrum discrete. 
The energies are \cite{Adam}: 
\be
\lb{En-nonrel}
E_n \ =\  \frac {\sigma z_k}\mu = z_k \left(\frac {\sigma^2}m \right)^{1/3} , \qquad n = 2k-1 = 1,3,5,\ldots,
 \ee 
where $-z_k$ is  the position of the $k$-th zero of the Airy function (see Fig. \ref{Airy-fig}). For large $k = 1,2,3,\ldots$,
\be
\lb{Airy-zero}
z_k \ =\  f_k^{2/3}\left(1 + \frac 5{48 f_k^2}  - \frac{5}{36 f_k^4} + \cdots  \right),
\ee
where 
 \be
f_k \ =\ \frac{3\pi (4k-1)}8.
 \ee

Keeping in \p{Airy-zero} only the leading term and substituting it in \p{En-nonrel}, we reproduce the semiclassical asymptotics 
\p{mu-nerel-WKB}.

For the even wave functions, $\Psi(0)$ does not vanish, but $\Psi'(0)$ does. Solving Eq. \p{Schr} with the Neumann boundary conditions at zero gives the spectrum
\be
\lb{En-nonrel1}
E_n \ =\   \frac {\sigma z'_k}\mu =  z'_k \left(\frac {\sigma^2}m \right)^{1/3} , \qquad n = 2k-2 = 0,2,4,\ldots
 \ee 
with
\be
\lb{Airy-crest}
z'_k  \ =\  g_k^{2/3}\left(1 - \frac 5{48 g_k^2}  + \frac{35}{288 g_k^4} - \cdots  \right),
\ee
where\footnote{Being expressed via $n$, both $f_k$ and $g_k$ boil down to the structure $\frac {3\pi}4 \left(n + \frac 12\right)$.}
 \be
g_k \ =\ \frac{3\pi (4k-3)}8.
 \ee
Now $-z'_k$ are  not the positions of the zeros but of the crests and troughs of the Airy function.

\section{Relativistic corrections}
\setcounter{equation}0

We now examine what happens if relativistic corrections to the spectrum are taken into account. In Ref. \cite{Ziya}, the 't Hooft equation was solved in the quasinonrelativistic regime where the ratios $\alpha_n = (M_n/2m)^2$ are close to one and may be represented as a series in the small parameter 
\be
\gamma \ =\  \left(\frac {g^2N}{4m^2} \right)^{2/3}\ = \   \left(\frac {\sigma}{m^2} \right)^{2/3}.
\ee

 The results for the odd and even levels are the following:
 \be
\lb{al-odd}
\alpha^{\rm odd}_n \ =\ 1 + \gamma z_k + \frac {\gamma^2 z_k^2}5 + \gamma^3 \left(- \frac 3{175} z_k^3 + \frac 6{35}  \right) +\cdots 
  \ee
and 
\be
\lb{al-even}
\alpha^{\rm even}_n = 1 + \gamma z'\_k + \gamma^2 \left(\frac { (z'\_k)^2}5 + \frac 1{5z'\_k} \right) + \gamma^3 \left(- \frac 3{175} (z'\_k)^3 + \frac 3{25}  - \frac 1{50 (z'\_k)^3}\right) +\cdots 
  \ee
with $z_k, z'_k$ defined in \p{Airy-zero} and \p{Airy-crest}.

We now evaluate $\alpha_n$ for the Hamiltonian \p{Salp} in the quasinonrelativistic region. The term $\propto \gamma$
corresponds to the nonrelativistic Schr\"odinger equation and is the same as in Eqs. \p{al-odd}, \p{al-even}. To evaluate the next term, one has to expand:
 \be
\hat H \ =\ 2m + \frac{\hat p^2}m - \frac {\hat p^4}{4m^3} + \cdots \ \equiv \ 2m + \hat H_0  - \frac {\hat p^4}{4m^3}  + \cdots 
 \ee 
The relativistic energy shift $E_n = M_n - 2m$ is
\be
\lb{shift}
\delta E_n  &\approx&  - \frac 1{4m^3} \langle \hat p^4 \rangle_n \ =\ - \frac 1{4m^3} \int_{-\infty}^\infty  \Psi_n(x) \Psi_n''''(x)  \, dx  = - \frac 1{4m^3} \int_{-\infty}^\infty  \Psi''_n(x) \Psi_n''(x) dx  \nn
& =&  - \frac 1{4m} \langle (E_n^{(0)} - \sigma |x|)^2 \rangle
\ =\  - \frac 1{4m}[ (E_n^{(0)})^2 -  2E_n^{(0)}\langle \sigma |x| \rangle_n + \sigma^2 \langle x^2 \rangle_n ], 
\ee
where  averaging is performed over the eigenstates of $\hat H_0$ and $E_n^{(0)}$ is the nonrelativistic energy.

\vspace{1mm}

An important comment is in order at this juncture. One might question the step $\int \Psi \Psi'''' dx \to \int \Psi'' \Psi'' dx$ that we used, saying that the boundary terms may also contribute something. But in fact they do {\it not}.

We begin with the odd states. Their wave functions have the form
\be
\lb{Psi-odd}
 \Psi_n(x) \ =\ {\rm sgn}(x) {\rm Ai}(\mu|x| - z_k)  \qquad {\rm with} \qquad \Psi_n(0) = {\rm Ai}(-z_k) = 0.
 \ee
One can verify that this function and also $\Psi_n'(x)$, $\Psi_n''(x)$ and $\Psi_n'''(x)$ are regular at $x=0$; only the fourth derivative is discontinuous. Then we may write:
\be
\lb{2int}
 \int_{-\infty}^\infty  \Psi_n(x) \Psi_n''''(x)  \, dx  =  \int_{-\infty}^0  \Psi_n(x) \Psi_n''''(x)  \, dx   + \int_0^\infty  \Psi_n(x) \Psi_n''''(x)  \, dx  .
 \ee
Flipping the derivative in the first integral, we derive:

 \be
\lb{boundary}
 \int_{-\infty}^0  \Psi_n(x) \Psi_n''''(x)  \, dx \ =\ -  \int_{-\infty}^0  \Psi'_n(x) \Psi_n'''(x) \, dx  + \Psi_n(0) \Psi'''_n(0).
 \ee
The boundary term is proportional to Ai$(-z_k)$ and vanishes. Flipping the derivative further in the remaining integral, we bring it in the form $\int \Psi'' \Psi'' dx$, while an extra boundary term is proportional to ${\rm Ai}'(-z_k) {\rm Ai}''(-z_k) =   {\rm Ai}'(-z_k) [-z_k {\rm Ai}(-z_k)]$ and vanishes too.  The same argument applies to the second integral in \p{2int}.

The wave function of an even state is 
\be
\lb{Psi-even}
\Psi_n(x) \ =\  {\rm Ai}(\mu|x| - z'_k)  \qquad {\rm with} \qquad \Psi'_n(0) = \pm \mu {\rm Ai}'(-z'_k) = 0.
 \ee
In this case, the boundary term in \p{boundary} is equal to $-\mu^3{\rm Ai}(-z'_k) {\rm Ai}'''(-z'_k)$, where we used 

$$ \Psi'''_n(-0)  =  \frac {\pd^3}{\pd x^3} {\rm Ai} (\mu|x| - z'_k) \left|_{x=-0}\ 
 \ =\  \frac {\pd^3}{\pd x^3} {\rm Ai} (-\mu x - z'_k) \right|_{x=0} = -\mu^3{\rm Ai}'''(-z'_k). $$

We have:

$${\rm Ai}'''(-z'_k) \ =\ \frac \pd{\pd v} {\rm Ai}''(v - z'_k) \left|_{v=0} \right. = \frac \pd{\pd v} [ (v - z'_k) {\rm Ai} (v - z'_k)]_{v=0} = {\rm Ai}(-z'_k) - z'_k {\rm Ai}'(-z'_k) = {\rm Ai} (- z'_k).$$

Thus, the boundary term is equal to $- \mu^3{\rm Ai}^2 (- z'_k)$ and does not vanish in this case. The second integral in \p{2int} gives the same contribution.

This is not all, however. For the function \p{Psi-even}, already the third derivative is discontinuous at zero, while the  fourth derivative has a singularity: $\Psi_n''''(x) \sim  2\mu^3 {\rm Ai} (-z'_k) \delta(x)$. The integral $\int_{-\infty}^\infty$ does not reduce in this case to the sum in the RHS of \p{2int}. The extra contribution due to this singular term in the LHS exactly cancels the boundary terms and the transformation in the first line of \p{shift} is thereby justified.

\vspace{1mm}

The virial theorem tells us:
\be
\langle U \rangle_n = 2\langle T \rangle_n \ \Longrightarrow  \ E_n^{(0)} = \langle U \rangle_n + \langle T \rangle_n
\ = \frac 32 \langle U \rangle_n  = \frac {3\sigma}2 \langle |x| \rangle_n.
 \ee
This gives 
\be
\lb{sdvig E}
\delta E_n \ \approx \ \frac 1{4m} \left[\frac { (E_n^{(0)})^2}3  - \sigma^2 \langle x^2 \rangle_n \right].
\ee
The calculation of $\langle x^2 \rangle_n$ is given in the Appendix. The result is:
\be
\lb{x2-av}
{\rm odd\ levels}: \qquad  \langle x^2 \rangle_n \ &=&\ \frac {8z_k^2}{15 \mu^2}, \nn
{\rm even\ levels}: \qquad  \langle x^2 \rangle_n \ &=& \ \frac 1{\mu^2} \left[\frac {8(z'_k)^2}{15 } + \frac 1{5z'_k} \right].
 \ee
Substituting this in \p{sdvig E}, we derive
\be
\delta E_n^{\rm odd} \ &=&\ - \frac {\sigma^2 z_k^2}{20 m\mu^2},  \nn
\delta E_n^{\rm even} \ &=&\ - \frac {\sigma^2}{20 m\mu^2} \left[ (z'_k)^2 + \frac 1{z'_k} \right].
\ee
 We therefore obtain:
 \be
\lb{al-Salp}
\alpha_n^{\rm odd}  &=& \frac {(2m + E_n^{\rm odd})^2}{4m^2} \ =\  1 + \gamma z_k + \gamma^2 \frac {z_k^2}5 + o(\gamma^2), \nn
\alpha_n^{\rm even}  &=& \frac {(2m + E_n^{\rm even})^2}{4m^2} \ =\ 1 + \gamma z'_k + \gamma^2 
\left[\frac {(z'_k)^2}5 - \frac 1{20 z'_k} \right]  + o(\gamma^2) 
 \ee

For the odd levels, this agrees with \p{al-odd},\footnote{It agrees at the level $\sim \gamma^2$, but not at higher levels. A numerical evaluation of the cubic contribution $\propto \gamma^3$ to $\alpha_1^{\rm odd}$ for the Salpeter Hamiltonian gives $C_3^{\rm Sal}(n=1) \approx -0.173$ whereas this coefficient in \p{al-odd} is equal to $C_3^{\rm Hooft}(n=1) \approx -0.048$.  } but for the even levels, we observe a disagreement with \p{al-even}.  In other words:

{\it The spectra evaluated by the Salpeter Hamiltonian \p{Salp} and by the 't Hooft Hamiltonian in \p{eq-Hooft} do not coincide once relativistic corrections are taken into account}.

However, note that the extra contributions $\propto \gamma^2/z'_k$ in \p{al-even} and in the second line of \p{al-Salp} are suppressed as $1/(z'_k)^3 \propto 1/n^2$ at large $n$, i.e. the disagreement only shows up in preasymptotic WKB corrections. In the leading semiclassical asymptotics, the spectra coincide, as they should.

\section{Numerical calculations}

\setcounter{equation}0

Some numerical results for the meson masses from the Bethe-Salpeter equation were displayed in the very first paper \cite{Hooft}, but no calculational details were given there. We use the discretization procedure outlined in Ref. \cite{Zubov} by R. Zubov and E. Prokhvatilov. These authors rewrote the 't Hooft equation in a more symmetric form, replacing the Hadamard finite-part integral with a principal-value integral:
 \be 
{\mathbcal m}^2 \phi(\rho) = \frac {4(\alpha +1)}{1-\rho^2} \phi(\rho) - 2\, {\rm  v.p.} \int_{-1}^1 \frac {\phi(\rho') - \phi(\rho)}{(\rho' - \rho)^2} d\rho',  
 \ee
where ${\mathbcal m}^2 = 2\pi M^2/(g^2N)$, $\alpha = 2\pi m^2/(g^2 N)$ and $\rho = 2r-1 \in (-1,1)$. Following \cite{Zubov}, we discretized the problem according to 
\be
&&\rho \to \rho_k  = \frac kK, \qquad k = -K, \ldots,-1,0,1, \ldots , K , \nn
&&\phi(\rho) \to \phi(\rho_k) \equiv \phi_k, \qquad \phi\_{-K} = \phi_K = 0, \nn
&&{\rm v.p.} \int_{-1}^1 \frac {\phi(\rho') - \phi(\rho)}{(\rho' - \rho)^2} d\rho' \ \to \ \frac 1K \sum_{k' = -K}^K \frac {\phi_{k'} - \phi_k }{(\rho_{k'} - \rho_k)^2}, \quad  k' \neq k
 \ee
with large enough $K$. Thus, the problem boils down to diagonalizing  a large matrix. This matrix can be diagonalized numerically.  

The calculation was performed at different $K = 400, 800, 1200, 1600$ and then the results were extrapolated to $K \to \infty$. The calculations proved sufficiently accurate.\footnote{Maybe it is due to the fact that the matrix elements decrease rapidly as we move away from the diagonal.} In particular, we were able to detect the difference between the analytical results of Ziyatdinov \p{al-odd}, \p{al-even} for the preasymptotic WKB corrections and the predictions \p{al-Salp} following from the relativistic Hamiltonian \p{Salp}.  The numerical results strongly favor \p{al-odd} and \p{al-even} and disfavor \p{al-Salp}.

To resolve this difference, we performed the calculations for comparatively large excitation numbers, $n=5$ and $n=6$, where the preasymptotic corrections are small enough so that one can keep only first few terms of the series. Then the difference between the predictions  \p{al-odd}, \p{al-even} and \p{al-Salp} shows up at the level of the fourth digit.
\begin{figure}[ht!]
 \begin{center}
        \subfigure[$n=0$]{
           \lb{Bor08}  \includegraphics[width=0.45\textwidth]{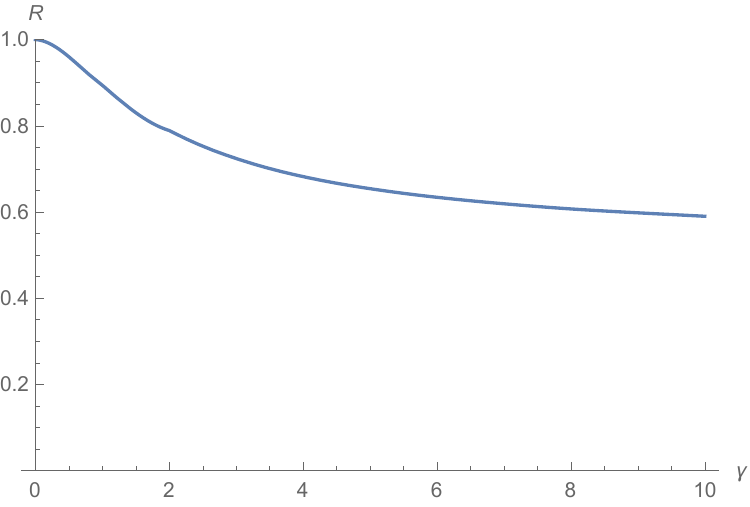}
        }
        \subfigure[$n=1$]{
         \lb{Bor11}
           \includegraphics[width=0.45\textwidth]{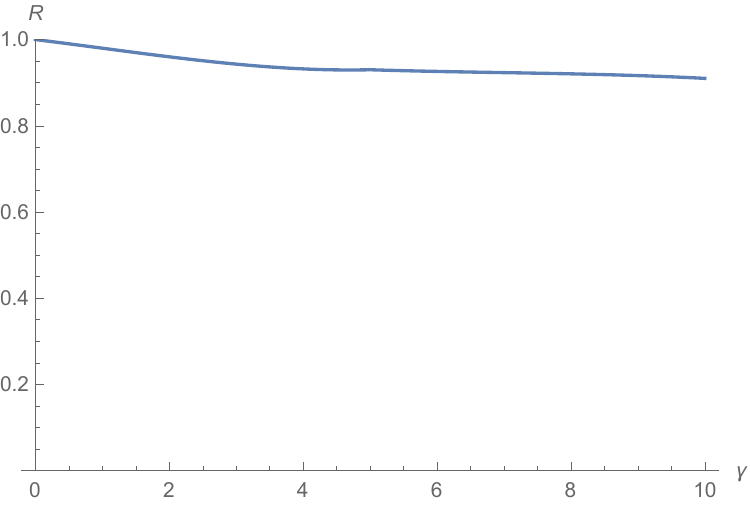}
        }
    \end{center}
    \caption{The ratio \p{ratio} as a function of
$\gamma = (\sigma/m^2)^{2/3}$ for $n=0$ and $n=1$ }
 \label{ratio-01}
    \end{figure}

The results for lower values of $n$ are also instructive. In Fig. \ref{ratio-01}, we represented the ratio
\be
\lb{ratio}
R \ =\ \frac{M^2_{\rm Salpeter}}{M^2_{\rm Hooft}}
 \ee
as a function of $\gamma$ for the ground state and the first excited state. (The spectrum of the Salpeter Hamiltonian was evaluated following the numerical procedure outlined in Sect. IIIB of Ref. \cite{NDD}.) We see that $R=1$ in the nonrelativistic limit; it then decreases and levels off at large $\gamma$. For $n=0$, it approaches  the value $R_\infty(n=0)$ somewhat above  one half. For the first excited state, this ratio is much closer to 1, never going down beyond $R = .9$.

\begin{figure}[ht!]
\begin{center}
 \includegraphics[width=0.45\textwidth]{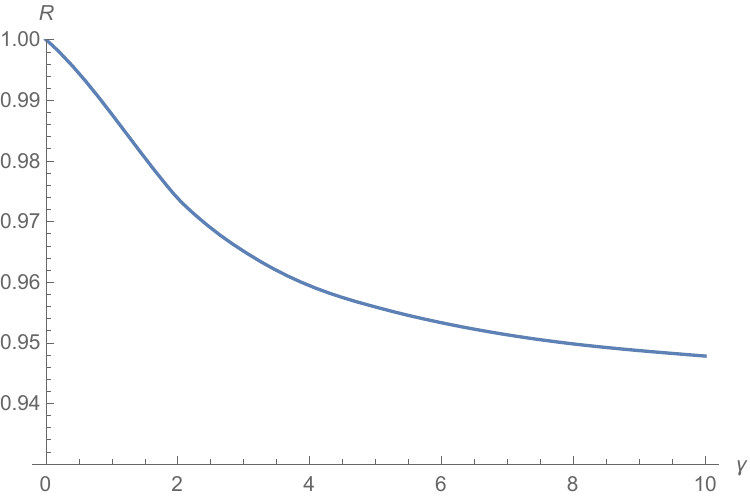}
           \end{center}
    \caption{The ratio \p{ratio} for $n=3$ }
\label{ratio3}
    \end{figure}

For $n=3$, the ratio remains close to 1 (see Fig. \ref{ratio3};  we changed the scale there  so that the bottom border of the graph is at $R= .9$). It is about $R \approx .95$ for $\gamma = 10$. The 5\% deviation for $M^2$ means just 2.5\% deviation for $M$. 

For still larger $n$ when the preasymptotic WKB corrections become less and less important, this ratio approaches to 1 also at large values of $\gamma$:
\be
\lb{R5-10}
R_{n=5}(\gamma = 10) \approx 0.964, \qquad   R_{n=10}(\gamma = 10) \approx .979 .
\ee

The deviations from unity can be readily understood by taking into account the leading preasymptotic corrections in the ultrarelativistic case. In relativistic quantum mechanics, Schr\"odinger equation is replaced by the Klein-Fock-Gordon equation. In the semiclassical analysis \cite{Bars1,Bars2,ja}, the action integral on the classical trajectory should be equated to $2\pi(n + 1/2)$ rather than $2\pi n$. In the ultrarelativistic case, this gives:

 \be
M^2_{\rm Salpeter}(n) \ =\ \frac {\pi g^2 N}2 \left(n + \frac 12 \right) + O\left( \frac 1n \right).
\ee 

For the 't Hooft equation, one derives instead\footnote{This was conjectured back in \cite{Hooft} and confirmed by a rather sophisticated analysis of the behavior of the solutions to the 't Hooft equation near the end points, $r \sim 0$ and $r \sim 1$, in Ref. \cite{Brower}.}

 \be
M^2_{\rm Hooft}(n) \ =\ \frac {\pi g^2 N}2 \left(n + \frac 34 \right) + O\left( \frac 1n \right).
\ee 

Consider the ratio of these two estimates:

$$
R^*_n(\infty) \ =\ \frac {n + 1/2}{n + 3/4} =  \frac {4n+2}{4n+3}.
$$

 We find that $R^*_{10}(\infty)  \approx 0.9767$ is very close to the numerical results in \p{R5-10}. 
For $\gamma = 50$ with $R_{n=10}(\gamma = 50) \approx 0.9763$, the agreement is even closer.

\vspace{1mm}

{\bf Conclusion.} The relativistic Hamiltonian and the 't Hooft equation give the same meson spectrum in two limits:
{\it  nonrelativistic limit, $g^2 \ll m^2$} and  {\it WKB limit}, $n \gg 1$.

In general, the spectra differ, although the difference is small from the second or third excited state onward. In all the values of parameters, the meson masses derived from the 't Hooft equation appear to be somewhat larger than the masses derived from the approximate Salpeter Hamiltonian.

\vspace{1mm}

I am grateful to Alexey Nefediev for illuminating discussions and for collaborating with me at the initial stage of this work.

\section*{Appendix: Calculation of $\langle x^2 \rangle_n$ for the Airy eigenfunctions}

\def\theequation{A.\arabic{equation}}

\setcounter{equation}0

We begin with the odd levels. We have:
 \be
\lb{A1}
\langle x^2 \rangle_n \ =\ \frac {\int_0^\infty x^2  {\rm Ai}^2\left(\mu x - z_k \right) dx}{\int_0^\infty  {\rm Ai}^2\left(\mu x - z_k \right) dx} \ = \ \frac {\int_{-z_k}^\infty (u + z_k)^2 {\rm Ai}^2(u) \, du }{ \mu^2\int_{-z_k}^\infty  {\rm Ai}^2(u) \,du}.
\ee 
We can write this as
\be
\lb{A2}
\langle x^2 \rangle_n \ =\ \frac 1{\mu^2} \left(z_k^2 + 2z_k \frac {I_1}{I_0} + \frac {I_2}{I_0}\right), \qquad {\rm with} \quad
I_p = \int_{-z_k}^\infty u^p {\rm Ai}^2(u)\, du.
\ee
We evaluate $I_{p=0,1,2}$. 
It follows from \p{Airy-eqn} that 
$$
{\rm Ai}^2(u) \ =\ \frac d{du} \left\{u {\rm Ai}^2(u) - [{\rm Ai}'(u)]^2 \right\}.
$$
Then 
\be
\lb{I0sum} 
I_0 =  [{\rm Ai}'(-z_k)]^2 + z_k {\rm Ai}^2(-z_k).
 \ee
We first consider the odd levels. In this case, ${\rm Ai}^2(-z_k) = 0$ and 
\be
\lb{I0}
I_0 \ =\ [{\rm Ai}'(-z_k)]^2.
\ee
To evaluate $I_1$, consider
 \be
\lb{av-x}
\langle x \rangle_n \ =\  \frac {\int_{-z_k}^\infty (u + z_k) {\rm Ai}^2(u) \, du }{ \mu \int_{-z_k}^\infty  {\rm Ai}^2(u) \,du} 
\ = \frac 1\mu \left( z_k + \frac{I_1}{I_0} \right).
\ee
On the other hand, using the virial theorem, we obtain:
$$
\langle U \rangle_n = \sigma \langle x \rangle_n = \frac {2E_n^{(0)}}3 = \frac {2\sigma z_k}{3\mu} \quad \Rightarrow \quad 
\langle x \rangle_n  = \frac{2z_k}{3\mu}.
$$
Comparing this with \p{av-x}, we derive
 \be
\lb{I1-I0}
\frac {I_1}{I_0} \ =\ - \frac {z_k}3.
 \ee
 The calculation of $I_2 = \int_{-z_k}^\infty u^2 {\rm Ai}^2(u)\, du $ is somewhat more tricky. Note first that
\be
\lb{for-I2}
u^2 {\rm Ai}^2(u) = u {\rm Ai}(u) {\rm Ai}''(u) \ = \  \frac d{du} \left[ u {\rm Ai}(u) {\rm Ai}'(u) - \frac 12 {\rm Ai}^2(u) \right] - u [{\rm Ai}'(u)]^2.
 \ee

The integral of the total derivative vanishes due to ${\rm Ai}(-z_k) = 0$, and we are left with

 \be 
\lb{I2-alt}
I_2 \ =\ -\int_{-z_k}^\infty u [ {\rm Ai}'(u)]^2 du.
  \ee
Consider now the identity 
\be
\lb{trick}
u^2 {\rm Ai}'(u) {\rm Ai}''(u) = u^3 {\rm Ai}(u) {\rm Ai}'(u) 
\ee
 and integrate it over $u$. Bearing in mind that 
$$u^2 {\rm Ai}'(u) {\rm Ai}''(u) \ =\  \frac 12  \frac d{du} \\{ u^2 [{\rm Ai}'(u)]^2\\} - u [{\rm Ai}'(u)]^2$$ 
and using \p{I0} and \p{I2-alt}, the integral of the left-hand side of \p{trick} may be represented as  $- z_k^2 I_0/2  + I_2$. 

And the integral of the right-hand side of \p{trick} is 
$$ 
\int_{-z_k}^\infty u^3 {\rm Ai}(u) {\rm Ai}'(u)\, du \ =\ \frac 12 (u^3 {\rm Ai}^2(u))  \Bigr|_{-z_k}^\infty   - \frac 32 \int_{-z_k}^\infty u^2 {\rm Ai}^2(u) \, du \ =\ - \frac 32 I_2.
$$
It follows that 
\be
\lb{I2}
- \frac {z_k^2}2 I_0 + I_2 = -\frac 32 I_2 \quad \Rightarrow \quad I_2 = \frac {z_k^2}5 I_0.
\ee 
Collecting all the terms, we are reproducing the first line in \p{x2-av}:
$$ 
\langle x^2 \rangle_n^{\rm odd} \ =\ \frac 1{\mu^2} \left(z_k^2 + 2 z_k \frac {I_1}{I_0} + \frac {I_2}{I_0} \right)\ =\ \frac 1{\mu^2} \left(z_k^2 - \frac{2z_k^2}3 + \frac{z_k^2}5 \right) \ =\ \frac  {8z_k^2}{15 \mu^2}.
 $$

\vspace{1mm}

We now turn to the even levels. Eqs. \p{A1}--\p{I0sum} retain the same form, with $z_k \to z'_k$, The boundary condition is now ${\rm Ai}'(-z'_k) = 0$, and this gives
 \be
I_0 \ =\ z'_k {\rm Ai}^2(-z'_k).
 \ee
Eq. \p{I1-I0} has also the same form with $z'_k$ substituted for $z_k$.

The calculation of $I_2$ follows the same steps as in the odd case, but the different boundary conditions introduce some modifications. In particular, the integral of the total derivative in \p{for-I2} does not vanish anymore, and we derive
\be
I_2 \ =\ \frac12 {\rm Ai}^2(-z'_k) - \int_{-z'_k}^\infty u [{\rm Ai}'(u)]^2.
 \ee
Take again the identity \p{trick} and integrate it. In the left-hand side, we obtain
\be
LHS \ &=&\ \int_{-z'_k}^\infty  u^2 {\rm Ai}'(u) {\rm Ai}''(u) \, du  \ =\ \frac 12 \left\{u^2 [{\rm Ai}'(u)]^2  \right\} \Bigr|_{-z'_k}^\infty  
- \int_{-z'_k}^\infty  u[ {\rm Ai}'(u)]^2\, du \nn
&=& \, - \int_{-z'_k}^\infty  u[ {\rm Ai}'(u)]^2\, du  = I_2 - \frac 12 {\rm Ai}^2(-z'_k).
 \ee

And in the right-hand side, we find 
\be
RHS  &=& \int_{-z'_k}^\infty  u^3 {\rm Ai}(u) {\rm Ai}'(u) \, du \ =\  \frac 12 \left[u^3 {\rm Ai}^2(u)  \right] \Bigr|_{-z'_k}^\infty  
\ - \frac 32 \int_{-z'_k}^\infty  u^2 {\rm Ai}^2(u)  \, du \nn
&=& - \frac 32 I_2 + \frac {z'^3_k}2 {\rm Ai}^2 (- z'_k).
\ee
Then
\be
\frac 52 I_2 \ =\ \frac 12 [{\rm Ai}'(-z'_k)]^2 [z'^3_k + 1) \quad \Rightarrow \quad I_2 = \frac {I_0}5 \left( z'^2_k + \frac 1{z'_k} \right).
\ee
We reproduce thereby the second line in \p{x2-av}:
$$ 
\langle x^2 \rangle_n^{\rm even} \ =\ \frac 1{\mu^2} \left(z'^2_k + 2 z'_k \frac {I_1}{I_0} + \frac {I_2}{I_0}\right) \ =\  \frac 1{\mu^2} \left[ z'^2_k - \frac{2z'^2_k}3 + \frac 15 \left({z'^2_k} + \frac 1{z'_k}\right) \right]\ =\ \frac 1{\mu^2} \left(\frac  {8z'^2_k}{15} + \frac 1{5 z'_k}  \right).
 $$

\end{document}